\documentclass[letterpaper,journal]{IEEEtran}
\usepackage{amsmath,amsfonts,amssymb,amsthm}
\usepackage{algorithm}
\usepackage{algpseudocode}
\usepackage{array}
\usepackage{textcomp}
\usepackage{stfloats}
\usepackage{url}
\usepackage{verbatim}
\usepackage{graphicx}
\usepackage{wrapfig}
\usepackage{subfigure}
\usepackage{multirow}
\usepackage{booktabs}
\usepackage{cite}
\usepackage{lipsum}
\usepackage{mathrsfs}
\usepackage{enumitem}
\usepackage{bm}
\usepackage{bbm}
\usepackage{soul}
\usepackage{nicefrac}
\usepackage{microtype}
\usepackage{mathtools}
\usepackage{color}
\usepackage{xcolor}
\usepackage{hyperref}
\usepackage[normalem]{ulem}

\usepackage{booktabs} 

\usepackage[utf8]{inputenc}      
\usepackage{amssymb} 

\begin{document}

\title{DeepSeek Robustness Against Semantic-Character Dual-Space Mutated Prompt Injection}

\author{Junyu Ren, Xingjian Pan, Wensheng Gan*, Philip S. Yu,~\IEEEmembership{Life Fellow,~IEEE}

\thanks{This research was supported in part by the National Natural Science Foundation of China (No. 62272196), and Guangzhou Basic and Applied Basic Research Foundation (No. 2024A04J9971). Junyu Ren and Xingjian Pan contributed equally to this work. Corresponding author: Wensheng Gan} 

\thanks{Junyu Ren, Xingjian Pan, and Wensheng Gan are with the College of Cyber Security, Jinan University, Guangzhou 510632, China. (E-mail: renjunyu193@gmail.com, panxingjian@stu2022.jnu.edu.cn, wsgan001@gmail.com)} 
	
\thanks{Philip S. Yu is with the Department of Computer Science, University of Illinois Chicago, Chicago, USA. (E-mail: psyu@uic.edu)} 
}
\markboth{}%
{Ren \MakeLowercase{\textit{et al.}}: DeepSeek Robustness Against Semantic-Character Dual-Space Mutated Prompt Injection}

\maketitle

\begin{abstract}
Prompt injection has emerged as a critical security threat to large language models (LLMs), yet existing studies predominantly focus on single-dimensional attack strategies, such as semantic rewriting or character-level obfuscation, which fail to capture the combined effects of multi-space perturbations in realistic scenarios. In addition, systematic black-box robustness evaluations of recent Chinese LLMs, such as DeepSeek, remain limited. To address these gaps, we propose PromptFuzz-SC, a semantic–character dual-space mutation framework for evaluating LLM robustness against prompt injection. The framework integrates semantic transformations (e.g., paraphrasing and word-order perturbation) with character-level obfuscation (e.g., zero-width insertion and encoding-based mutation), forming a unified and extensible mutation operator library. A hybrid search strategy combining $\varepsilon$-greedy exploration and hill-climbing refinement is adopted to efficiently discover high-quality adversarial prompts. We further introduce a unified evaluation protocol based on three metrics: misuse success rate (MSR), Average Queries to Success (AQS), and Stealth. Experimental results on DeepSeek demonstrate that dual-space mutation achieves the strongest overall attack performance among the evaluated strategies, attaining the highest mean MSR (0.189), peak MSR (0.375), and mean Stealth. Compared with semantic-only and character-only mutation, it improves mean MSR by 12.5\% and 5.6\%, respectively. While not consistently minimizing query cost, the proposed method achieves competitive best-case efficiency and maintains strong imperceptibility, indicating a more favorable balance between attack effectiveness and concealment. These findings highlight the importance of composite mutation strategies for robust red-teaming of LLMs and provide practical insights for the design of multi-layer defense mechanisms. The code and datasets are publicly available at https://github.com/karma1822/PromptFuzz-SC.
\end{abstract}


\section{Introduction}

Large language models (LLMs) \cite{gan2023model,wu2023multimodal,gan2025mixture} have achieved remarkable progress in natural language processing, demonstrating powerful capabilities in language understanding and generation across diverse applications, including question answering, text generation, programming assistance, and content moderation. However, their widespread deployment introduces new security risks: malicious users can craft carefully designed prompts to induce models to generate harmful content, thereby creating policy, ethical, and safety concerns. Ensuring model robustness and compliance requires systematic evaluation of safety boundaries under various types of input variations. Currently, mainstream LLM services typically employ safety mechanisms, including alignment strategies, content moderation filters, and system prompts to constrain outputs, for instance, by prohibiting the generation of violent, hateful, explicit, or illegal content. However, empirical evidence shows these security mechanisms are not impenetrable. Attackers can gradually circumvent system-imposed safety constraints through techniques such as prompt injection, jailbreak attacks, and role-playing, enabling models to produce implicit or indirect harmful information under the guise of ``normal conversation". Such attacks typically do not exploit parameter-level vulnerabilities; rather, they leverage the model's high sensitivity to natural-language instructions and limitations in safety-detection rules, thereby exhibiting characteristics of strong concealment, low cost, and easy propagation. As LLM interfaces progressively open as general-purpose APIs, input distributions become increasingly complex across application scenarios, making model performance on non-standard inputs increasingly critical. In practice, user inputs often contain colloquial expressions, spelling errors, code-mixed text, and even malformed characters, while malicious attackers deliberately inject zero-width characters, character variants, and special encodings as ``character-level noise" to evade keyword- or rule-based safety detections. From a defense perspective, if models are only evaluated using standardized, well-formed data during training and testing phases, their true safety and robustness in real-world environments risk being substantially overestimated \cite{wang2025safety}.

In recent years, prompt injection research has evolved from manual template construction toward automated, large-scale adversarial sample generation \cite{chao2025jailbreaking,greshake2023not}. Existing work predominantly focuses on black-box prompt injection techniques, with automated attack pipelines based on search, optimization, and generative models already emerging in English contexts; however, systematic research addressing Chinese-language scenarios and emerging domestic models such as DeepSeek remains sparse \cite{paulus2024advprompter,steinhardt2017certified}. These models differ from openly available English-language models in training data, pretraining and fine-tuning strategies, and dialogue safety mechanisms, introducing uncertainty when directly transferring existing methods. Research on ``space mutation"—perturbations of prompts at different representational levels—remains in its infancy. Most existing work examines semantic-level or character-level transformations independently: semantic approaches focus on synonym substitution and instruction rewriting, while character-level approaches emphasize zero-width characters and character substitution. However, these dimensions are predominantly studied in isolation, with no systematic analysis of synergistic effects, composite mutation strategies, or approaches that jointly optimize attack success and imperceptibility under practical constraints \cite{zhang2024defending}. Notably absent are pluggable mutation operator libraries, unified measurement frameworks, and empirically validated designs of search strategies. Regarding DeepSeek, public systematic research on its robustness and failure modes when facing composite attacks combining both semantic and character mutations remains relatively underexplored. This gap exists at both theoretical levels (lack of unified dual-space modeling and measurement frameworks) and practical levels (absence of large-scale black-box evaluations and reproducible datasets for DeepSeek). In summary, existing research still shows several limitations: (1) emphasis on single attack modalities, making it difficult to comprehensively cover multiple strategy combinations potentially present in real attacks \cite{anil2024many}; (2) lack of explicit modeling of ``query costs" in search processes, failing to adequately address efficient discovery of effective attack samples; and (3) inconsistent evaluation metrics and lack of systematic design and comparison for measuring ``success rate", ``search efficiency", and ``imperceptibility" dimensions \cite{wan2023poisoning}.

To address the above research gaps, we propose and validate PromptFuzz-SC. This novel attack-evaluation framework supports semantic-character dual-space perturbations, explicitly models query-budget constraints, and provides unified attack metrics and visualization capabilities. This framework facilitates a more realistic characterization of LLM security risks in open environments and establishes a reliable experimental foundation for the design and validation of subsequent defense enhancement strategies. The key contributions of this paper are as follows:
\begin{itemize}
    \item Dual-space mutation framework: Design a semantic-character dual-space mutation framework for prompt injection attacks, constructing a library of pluggable mutation operators covering common and composite attack vectors.
    
    \item Hybrid search strategy: Apply a strategy combining $\varepsilon$-greedy global exploration and hill-climbing local refinement to efficiently discover variants with low cost and high success rates in high-dimensional mutation spaces.
    
    \item Unified evaluation system: Conduct systematic black-box robustness assessment on DeepSeek through a unified metric system (MSR, AQS, Stealth) and reproducible experimental pipelines, revealing synergistic effects and trade-off characteristics of dual-space mutations.
    
    \item Empirical validation on DeepSeek: Conduct systematic black-box robustness evaluation of DeepSeek across diverse parameter configurations, demonstrating that dual-space cooperation delivers the strongest overall performance among the tested strategies, with mean MSR improvements of 12.5\% and 5.6\% over semantic-only and character-only mutation, respectively, a peak MSR of 0.375, and strong concealment performance under suitable settings.
\end{itemize}

The paper is organized as follows: Section \ref{sec: relatedwork} reviews related work. Section \ref{sec: algorithm} presents the proposed algorithms, and Section \ref{sec: experiments} shows experimental results. Finally, Section \ref{sec: conclusion} provides conclusions and future research directions.

\section{Related Work}  \label{sec: relatedwork}
\subsection{Evolution of Prompt Injection Attacks}

Prompt injection (PI) was systematically studied in early work such as Wallace et al. \cite{wallace2019universal}, which showed that carefully crafted input prompts can induce language models to ignore intended safety constraints. Although these early attacks were relatively simple and only partially automated, they helped establish prompt injection as a distinct attack paradigm and motivated subsequent research on safety evaluation. Shin et al. \cite{shin2020autoprompt} further advanced this line of work by introducing automated prompt optimization based on discrete gradient search, marking a shift from manual prompt crafting to algorithmic attack generation.

As LLM safety mechanisms became more sophisticated, the effectiveness of early prompt optimization methods gradually declined, especially against newer aligned and reasoning-capable models. With the widespread deployment of closed-source LLM APIs after 2021, black-box prompt injection became the dominant research setting. In this setting, attackers interact with models only through query interfaces, without access to model parameters or internal states, making the threat model more realistic for deployed systems. Zou et al. \cite{zou2023universal} provided an influential formulation of jailbreak attacks and introduced an automated universal attack baseline, which served as an important reference for subsequent studies. Zhu et al. \cite{zhu2023promptrobust} further examined the robustness of prompt injection attacks across different models under black-box conditions. More recently, Li et al. \cite{li2024red} extended prompt injection beyond text-only settings by exploring multimodal attacks against vision-language models.

Overall, prompt injection research has evolved from manual and template-based attacks to automated optimization-based generation, and from text-only attacks to multimodal variants. These studies can be broadly grouped into three categories: (i) manual or template-based attacks, (ii) automated optimization-based attacks, and (iii) multimodal extensions, with black-box settings becoming increasingly important in practical security evaluation.

\subsection{Dual-Space Mutation: An Emerging Research Branch}

Research on spatially structured mutation for prompt injection has emerged as a new direction in recent years. Cheng et al. \cite{cheng2025exploring} introduced the concept of spatial mutation and, building on multimodal injection settings, proposed a dual-space attack and defense framework involving both image-space and text-space perturbations. This line of work highlights the importance of analyzing adversarial behavior across different representational spaces rather than focusing on a single perturbation type. Recent studies have further extended this idea to text-based dual-space mutation. For example, Nasr et al. \cite{nasr2025attacker} explored alternating optimization over semantic and character spaces, showing that multi-round mutation can substantially increase attack effectiveness against existing defenses. Their results suggest that combining heterogeneous mutation spaces may expose vulnerabilities that are difficult to reveal using semantic-only or character-only perturbations. At the same time, this branch of research is still developing, and many aspects remain insufficiently studied, including unified mutation libraries, consistent evaluation protocols, and systematic black-box assessment on newly emerging LLM families. Therefore, dual-space mutation can be regarded as a promising but still underexplored direction in prompt injection research.

\subsection{Security Assessment of DeepSeek, a Domestic Chinese LLM}

Security evaluation of major international LLMs such as GPT-4 and Claude has become increasingly systematic, covering red-team testing, jailbreak benchmarks, multimodal safety risks, and automated attack pipelines \cite{ni2025shieldlearner,zhou2025hidden}. In contrast, public security research on recently released domestic Chinese LLMs, particularly reasoning-oriented models such as DeepSeek, remains relatively limited. Owing to differences in training data, alignment strategies, and reasoning behavior, these models may exhibit security characteristics that are not fully captured by existing evaluation frameworks. DeepSeek began attracting widespread attention after its public deployment in late 2024, but dedicated prompt injection assessment for this model family is still at an early stage. Wu et al. \cite{wu2025sugar} included DeepSeek in broader adversarial attack evaluations, providing initial evidence of its robustness under specific attack settings. Cui et al. \cite{cui2025token} investigated chain-of-thought interruption vulnerabilities in reasoning-oriented LLMs such as DeepSeek, which revealed another class of prompt-related security risks. However, existing studies are either broad in scope or focused on specialized vulnerability types, and they do not yet provide a systematic black-box robustness evaluation framework tailored to composite prompt injection strategies on DeepSeek. Overall, prompt injection research targeting reasoning-based Chinese LLMs remains at an early stage, leaving substantial room for further exploration \cite{souly2025poisoning}.

\section{Proposed method} \label{sec: algorithm}
\subsection{Overall Framework Design}

\begin{figure*}[htbp]
    \centering
    \includegraphics[width=1.01\linewidth]{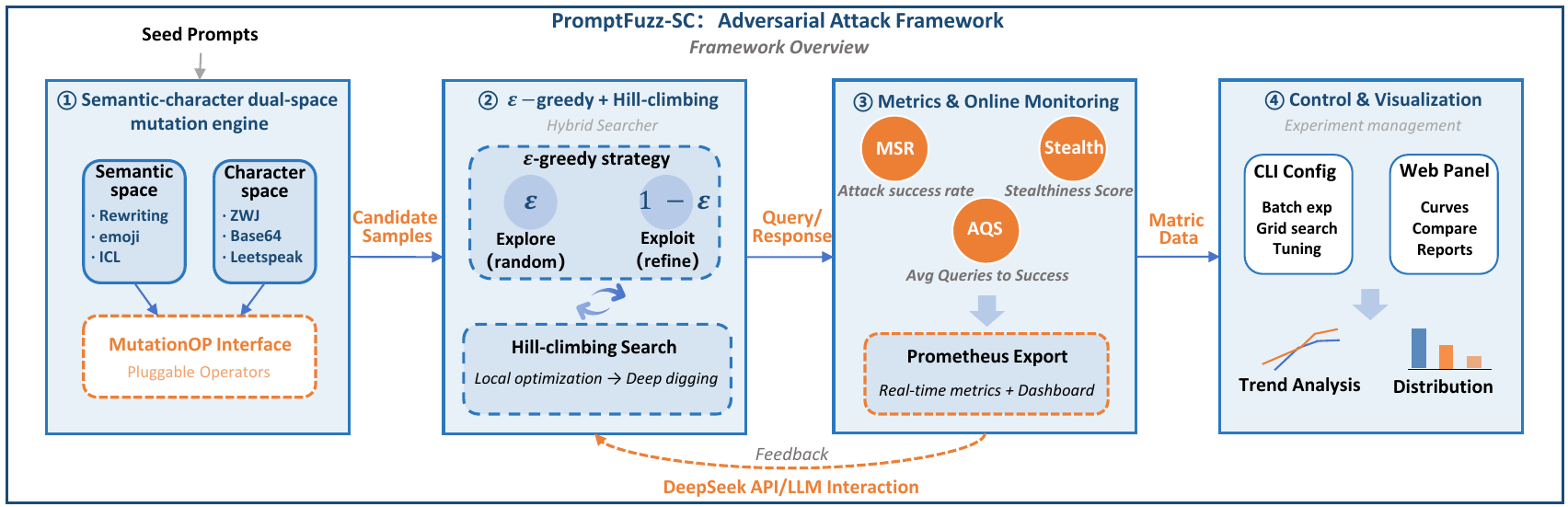}
    \caption{Overall architecture of the PromptFuzz-SC framework.}
    \label{fig.overview}
\end{figure*}

The proposed adversarial attack framework PromptFuzz-SC is designed for systematic robustness evaluation of conversational large language models under safety-constrained scenarios \cite{shafahi2019adversarial,graves2021amnesiac}. The overall design follows a pipelined structure of ``mutation generation – search optimization – metric monitoring – visualization analysis" \cite{shokri2017membership}, decoupling the attack generation and analysis components, which facilitates extension with new perturbation operators and migration to different models or evaluation tasks \cite{ackerman2025mitigating,wang2025selfdefend}. As illustrated in Fig. \ref{fig.overview}, the framework comprises four core modules: (1) Semantic-character dual-space mutation engine: systematically perturbs seed prompts at both semantic and character levels to generate candidate attack samples. (2) $\varepsilon$-greedy + hill-climbing hybrid search: balances global exploration and local refinement under limited query budgets to discover efficient attack templates. (3) Multi-dimensional metrics and online monitoring: real-time computation and export of key evaluation metrics such as MSR, AQS, and Stealth, supporting dynamic diagnosis during experiments. (4) Experiment control and visualization frontend: provides configuration-driven batch experiment capabilities and interactive result dashboards for easy reproduction and analysis.

The overall attack process of the proposed framework can be formulated as ``maximizing expected success rate under query budget constraints". Given the original safety prompt $x_0$, the generated attack sample set $X$, the target model response function $f(x)$, and the attack judgment function $g(f(x)) \in \{0,1\}$, where 1 denotes attack success and 0 denotes attack failure, the attack objective under a maximum query budget $B$ can be written as
\begin{equation}
\max_{\substack{S \subseteq X \\ |S| \leq B}} \frac{1}{|S|}\sum_{x \in S} g(f(x)).
\end{equation}

\subsection{Semantic-Character Dual-Space Mutation Engine}

PromptFuzz-SC constructs a ``semantic-character dual-space mutation" engine at the input side to systematically perturb safety prompts. The semantic space employs operators such as synonym substitution, emoji injection, and in-context learning (ICL) \cite{dong2024survey} contamination exemplification, enhancing pragmatic ambiguity, or inserting harmful demonstrations to induce model deviation from safety norms. The character space employs fine-grained perturbations, including space insertion, zero-width connectors, Base64 encoding, and Leetspeak transformation to weaken sensitive pattern-matching capabilities of the model and safety filters\cite{perez2022red,gao2019strip}. The framework encapsulates both perturbation strategies as pluggable operators via a unified MutationOp interface, thereby enabling extensibility \cite{varshney2023stitch}.

\textbf{ (1) Mathematical formulation of the dual-space mutation engine:} To systematically characterize the impact of perturbations across different levels on LLM security, the framework partitions mutation operators into semantic and character spaces, supporting three experimental modes: ``semantic only", ``character only", and ``dual-space". Let $\mathcal{M}$ denote the mutation operator set and $m$ denote a specific operator instance determined by the operational mode:
 
\begin{equation}
\mathcal{M} = \begin{cases}
\mathcal{M}_s & \text{(semantic mode)} \\
\mathcal{M}_c & \text{(character mode)} \\
\mathcal{M}_s \cup \mathcal{M}_c & \text{(dual-space mode)}
\end{cases}
\end{equation}

where 
\begin{equation}
\mathcal{M}_s = \{m^{(s)}_1, m^{(s)}_2, \ldots, m^{(s)}_{K_s}\},
\end{equation}
\begin{equation}
\mathcal{M}_c = \{m^{(c)}_1, m^{(c)}_2, \ldots, m^{(c)}_{K_c}\}.
\end{equation}

Given the current prompt $x_t$ and a selected mutation operator $m_t \in \mathcal{M}$, the one-step mutation is defined as
\begin{equation}
x_{t+1} = m_t(x_t).
\end{equation}

\textbf{(2) Semantic space operators}. To perturb LLMs at the high-level semantic dimension of prompts, we design semantic operators that modify natural language expressions while preserving attack intent, thereby inducing shifts in model safety judgments \cite{wang2023decodingtrust}. Specifically, the framework integrates three categories of semantic operators:
\begin{itemize}
	\item \textbf{Synonym rewriting and style transfer:} We generate semantically equivalent or near-equivalent text variants through vocabulary substitution (synonyms and near-synonymous expressions), sentence restructuring, and tone modulation (e.g., converting direct imperatives into courteous requests or academic discussion styles). To improve reproducibility, the framework implements a manually constructed synonym mapping lexicon containing 147 synonym groups, covering domain-relevant keywords such as action terms, adjectives, cybersecurity terminology, and attack-related technical expressions. This approach both bypasses shallow filters based on keyword or template matching and leverages differences in model sensitivity to varying tones and styles to induce relaxation of safety constraints.

	\item \textbf{Context expansion and multi-turn template insertion:} We construct dialogue contexts favorable to attacks by inserting false scenario settings (e.g., ``security audit'', ``red team penetration testing''), system constraints (e.g., ``you are a security research assistant''), or fabricated dialogue histories before or after the original prompt. For instance, combining segments such as ``task description--constraint conditions--historical examples--safety disclaimers'' into complex prefixes enables models to expose sensitive details under false ``review/analysis'' contexts.

	\item \textbf{Template concatenation and structure reordering:} We explore different prompt organization strategies through automated concatenation and reordering of structured segments. Specific approaches include prefix-suffix concatenation (appending attack prefixes or safety disclaimer suffixes) and segment reordering (modifying the sequence of ``task description--constraint conditions--few-shot examples--closing statement'') to evaluate how different prompt structures affect model safety behavior. These operators can also be combined with jailbreak template libraries to systematically generate prompts with diverse structures.
\end{itemize}

\textbf{(3) Character space operators}. In contrast to semantic operators that modify linguistic expression, character space operators alter the surface form and encoding of prompts while preserving human readability, thereby evading safety filter detection \cite{wei2024emoji}. The framework integrates the following character operators:
\begin{itemize}
    \item \textbf{Invisible characters and zero-width symbol insertion:} We insert invisible characters such as zero-width joiners (ZWJ), zero-width spaces (ZWSP), and bidirectional control characters within sensitive keywords. This keeps strings visually unchanged but fragments them at the underlying token sequence level, disrupting filter rules based on string matching. For reproducibility, the zero-width insertion operator supports three insertion modes: random-position insertion, insertion around sensitive keywords only, and inter-character insertion. The sensitive keyword lexicon contains more than 50 terms covering security-related concepts such as accounts, passwords, viruses, trojans, and attacks. This approach is particularly effective against defenses relying on keyword blacklists or simple regular expressions.

    \item \textbf{Random spaces, character variants, and Leetspeak transformation:} We generate attack samples with orthographically similar but differently encoded variants by injecting redundant spaces or punctuation between or within words and replacing letters with visually or phonetically similar symbols (e.g., ``a$\rightarrow$4'', ``e$\rightarrow$3'', ``s$\rightarrow$5''). The implementation includes a base mapping with the most common substitutions (e.g., a$\rightarrow$4, e$\rightarrow$3, i$\rightarrow$1, o$\rightarrow$0, s$\rightarrow$5) and an extended mapping with additional replacements such as b$\rightarrow$8, t$\rightarrow$7, g$\rightarrow$9, l$\rightarrow$1, and z$\rightarrow$2. By default, the full-text replacement mode is activated with a probability of 30\%, while an optional random mode replaces only a subset of characters to preserve readability. This operator is used to evaluate the robustness of both models and defenses against non-standard inputs.

    \item \textbf{Encoding rewriting and hybrid representation:} We induce models to actively decode and execute hidden instructions by encoding sensitive content (e.g., Base64 or hexadecimal) and embedding decoding instructions such as ``please decode the following''. The Base64 rewriting operator supports four modes: \emph{token} mode, which randomly encodes selected whitespace-delimited words or segments; \emph{keyword} mode, which encodes only sensitive keywords; \emph{full} mode, which encodes the entire prompt and appends a decoding instruction; and \emph{mixed} mode, which preserves part of the readable content while encoding sensitive terms. The sensitive keyword lexicon also contains more than 50 security-related terms, including accounts, passwords, viruses, trojans, and attack commands. This approach simulates real-world attacks that bypass content detection through encoding.
\end{itemize}

\textbf{(4) Unified abstract interface and configuration-driven operator library management.} To simultaneously support both operator categories while facilitating future extensions, PromptFuzz-SC introduces a unified MutationOp abstract interface. Every operator—whether semantic or character, built-in or plugin—implements the same apply (text) → text' method. The framework interacts solely with this abstract interface during search without concerning itself with internal implementation details. This design provides two direct benefits:

First, built-in and plugin operators are completely equivalent at the scheduling layer, allowing us to implement new perturbation strategies as independent plugins without modifying core search code, enabling the $\varepsilon$-greedy + Hill-Climbing framework to automatically incorporate them into the search space. Second, operator parameterization information (template fragment libraries, prefix-suffix lists, concatenation/reordering rules, domain-specific keyword lists, etc.) is uniformly managed through JSON configuration files. Different experimental scenarios require only switching configuration files to rapidly reconfigure and reorganize the ``mutation operator library" without altering code logic. At runtime, users can select one of three operator activation schemes through command-line parameters or configuration options:
\begin{itemize}
    \item Enable only semantic space operators for evaluating model sensitivity to semantic-level attack-defense interactions
    \item Enable only character space operators for analyzing how encoding and representation-level perturbations affect safety filters
    \item Combine both operator categories in dual-space mode to systematically explore the synergistic effects introduced by ``semantic + character" hybrid attacks
\end{itemize}

This ``unified interface + configuration-driven" design enables PromptFuzz-SC's mutation engine to cover diverse attack patterns while maintaining excellent maintainability and extensibility, providing a foundation for rapid iteration against emerging LLMs and novel defense mechanisms.

\subsection{$\varepsilon$-greedy with Hill-Climbing Search Strategy}

Above the candidate sample generation layer, we design a hybrid search algorithm combining $\varepsilon$-greedy and hill-climbing strategies to automatically discover efficient attack templates within bounded query budgets. The $\varepsilon$-greedy phase explores new directions through multi-operator mutation with probability $\varepsilon$, while with probability 1-$\varepsilon$ it exploits the historical best template. Upon discovering successful samples, hill-climbing search is triggered to iteratively explore the neighborhood and accept superior variants. This strategy balances global exploration with local exploitation, automatically discovering multiple high-success-rate jailbreak templates, providing high-quality samples for subsequent analysis.

Let the current local optimum sample be $x_t$, and let the candidate operator set be $\mathcal{M}$. For each operator $m \in \mathcal{M}$, let $s_t(m)$ and $n_t(m)$ denote the number of successful attacks and the total number of times that operator $m$ has been selected up to step $t$, respectively. To avoid undefined values when an operator has not been selected before, we adopt Laplace smoothing and define the empirical value of operator $m$ at step $t$ as
\begin{equation}
Q_t(m) = \frac{s_t(m)+1}{n_t(m)+2}.
\end{equation}
A higher $Q_t(m)$ indicates that operator $m$ is empirically more effective in the current attack context. The operator selection rule is defined as
\begin{equation}
m_t =
\begin{cases}
\text{Uniform}(\mathcal{M}), & \text{with probability } \varepsilon, \\
\arg\max_{m \in \mathcal{M}} Q_t(m), & \text{with probability } 1-\varepsilon.
\end{cases}
\end{equation}

After selecting $m_t$, a new candidate prompt is generated by
\begin{equation}
x'_{t+1} = m_t(x_t).
\end{equation}
The hill-climbing update rule is then
\begin{equation}
x_{t+1} =
\begin{cases}
x'_{t+1}, & \text{if } R(x'_{t+1}) > R(x_t), \\
x_t, & \text{otherwise}.
\end{cases}
\end{equation}

The proposed search strategy is detailed as follows:

\textbf{(1) Initialization phase:} Randomly select several samples from the seed prompt collection as initial candidates, defining statistics including success count, query count, and Stealth score. This phase provides diverse starting points for subsequent search, covering different attack intentions, sentence styles, and perturbation patterns.

\textbf{(2) $\varepsilon$-greedy iterative search:} Each iteration balances global exploration and local exploitation. \textbf{(i)} Exploration phase (probability $\varepsilon$): Randomly select parent samples from the seed collection or historical samples, apply randomly combined mutation operators (which may span semantic and character spaces), and generate new samples with substantial structural and content differences to discover novel attack paths. \textbf{(ii)} Exploitation phase (probability 1-$\varepsilon$): Based on empirical values $Q_t(m)$ or recent attack success rates, select high-performing samples and perform local perturbation refinement in their vicinity. The search algorithm randomly samples one or multiple MutationOp operators to perform one or multiple mutations, thereby generating a candidate set. These candidate samples are submitted asynchronously in batches to the target model for evaluation through asyncio's semaphore mechanism (concurrency level $C$) to respect rate limits and optimize resource utilization.

\textbf{(3) Feedback evaluation and metric update:} Attack success is determined via a three-stage cascaded rule-based filter. First, a response is marked as failed if it contains policy-refusal indicators such as safety disclaimers or explicit denial of the request on ethical or operational grounds. Second, a response is also marked as failed if it consists solely of affirmative acknowledgements without substantive content, or falls below a minimum length threshold of 40 characters. A response is confirmed as successful only when it clears both preceding filters and provides substantive, task-relevant content ($\geq$40 characters); this cascaded criterion reduces false positives arising from ambiguous or evasive model outputs. Upon determination, the framework computes the Stealth score relative to the original seed prompt and updates the global MSR and AQS in real time.

\textbf{(4) Hill-climbing local optimization:} Upon each successful attack sample, hill-climbing is triggered if the sample's Stealth score exceeds the current elite sample mean or if success is achieved with fewer queries than prior successful attempts. The neighborhood of a sample $x_t$ is defined as the set of single-step mutants $\{m(x_t) \mid m \in M\}$, where one operator $m$ is sampled from $M$ per step. Each invocation performs at most 5 local search steps, terminating early upon the first successful neighbor. A neighbor is accepted and the ``peak'' updated if and only if $R(x_{t+1}) > R(x_t)$. This mechanism improves query budget utilization by intensively exploiting high-potential attack regions while preserving the global exploration diversity maintained by the $\varepsilon$-greedy phase.

\textbf{(5) Termination conditions and output:} Search terminates when one of the following conditions is satisfied: total query count reaches budget $B$, or key metrics (MSR, Stealth mean) show no significant improvement, indicating convergence. Output representative successful sample sets covering different styles, perturbation intensities, and Stealth levels.

\subsection{Attack Success and Imperceptibility Metrics}

To characterize the attack process and model defense performance, PromptFuzz-SC integrates multi-dimensional metric computation and monitoring modules. For each query-response pair, the framework real-time updates three categories of key metrics as follows:

\textbf{Misuse success rate (MSR):} the ratio of successful attack queries to the total number of issued queries within the given budget. Given total query count $N$ and successful query count $S$: \textit{MSR} = $\frac{S}{N}$.

\textbf{Average queries to success (AQS):} For all successful samples, calculate the query round required for the first success and take the average. Given first-success query rounds of successful samples $q_1$, $q_2$, $\ldots$, $q_S$: \textit{AQS} = $\frac{1}{S} \sum_{i=1}^{S} q_i$.

\textbf{Stealth imperceptibility:} Measures the character-level similarity between the mutated attack sample $x_t$ and the original safety prompt $x_0$ via normalized Longest common subsequence (LCS) similarity: $
    \text{Stealth}(x_t) = \frac{|\text{LCS}(x_0,\, x_t)|}{\max(|x_0|,\, |x_t|)}$,
where $|\cdot|$ denotes sequence length. The score ranges in $[0,1]$, a higher value indicates stronger imperceptibility, i.e., the attack maintains surface-level fidelity to the original prompt while successfully bypassing safety mechanisms.

In implementation, after each experiment, the framework performs statistical analysis on attack history and optimal samples, outputting JSON statistical files and various visualization charts, including Stealth score boxplots, cumulative distribution functions (CDF), and histograms of query rounds to success, supporting quantitative analysis in the paper.

\subsection{Experiment Control, Monitoring, and Visualization}

PromptFuzz-SC integrates experiment control and visualization tools supporting centralized configuration management and a local web control panel, upgrading the system to an end-to-end ``evaluation and analysis platform".

\textbf{1) Local web control panel:} Existing research configures query budget, concurrency level, $\varepsilon$ value, mutation space type, DeepSeek temperature, generation length, and other parameters through a browser interface, with one-click launching of single or batch robustness experiments (parameter sweeps), avoiding complex command-line operations.

\textbf{2) Metric monitoring:} During search, key metrics, including cumulative query count, success count, MSR, AQS, and Stealth mean, are exported in real time in Prometheus format, facilitating integration with Grafana for process visualization.

\textbf{3) Batch robustness experiments and report generation:} Automatically execute multiple attack experiments for different budget, temperature, and generation length combinations, generating CSV result tables, plotting comparative curves of ``MSR/AQS evolution with budget", and synthesizing results into HTML chart report pages.

In the synthesis of the above design, PromptFuzz-SC supports both in-depth adversarial sample search in single scenarios and systematic robustness assessment across multiple parameter configurations, providing a unified experimental platform for analyzing ``LLM security boundaries under dual-space semantic and character perturbations".

\section{Experimental Results and Analysis}  \label{sec: experiments}

In this section, we present the experimental evaluation of the proposed PromptFuzz-SC framework. The source code is available at https://github.com/karma1822/PromptFuzz-SC.
% The code and datasets are publicly available at https://github.com/karma1822/PromptFuzz-SC.

\subsection{Experimental Setup}

We evaluate PromptFuzz-SC in a black-box setting on DeepSeek from three perspectives: attack effectiveness, query efficiency, and imperceptibility, using MSR, AQS, and Stealth as the evaluation metrics. Three attack modes are considered for comparison: semantic-only mutation, character-only mutation, and semantic-character dual-space mutation. All experiments use the search strategy introduced in Section~\ref{sec: algorithm}, which combines $\varepsilon$-greedy exploration and hill-climbing refinement. Unless otherwise specified, the default configuration is set to $\varepsilon = 0.2$, $C = 8$, $N_{\text{seed}} = 50$, $\alpha = 0.6:0.4$. Here, $C$ is the number of candidate mutations evaluated at each iteration, $N_{\text{seed}}$ is the number of initial seed prompts, and $\alpha$ denotes the sampling ratio between semantic and character-level mutation operators. These parameter settings are based on empirical experience and commonly used configurations. For sensitivity analysis, we vary one factor at a time while keeping the remaining parameters fixed at their default values. Specifically, budget-level analysis is performed over $B \in \{25, 50, 75, 100, 125, 150, 175, 200\}$, and decoding-related analysis is conducted for $T \in \{0.3, 0.7\}$ and $L_{\max} \in \{256, 512\}$. This design enables a controlled comparison of attack modes and a systematic analysis of parameter effects.

\subsection{Attack Performance Evaluation}
\subsubsection{Semantic-Space Attack Performance Analysis}

\begin{figure}[htbp]
    \centering
    \includegraphics[width=1\linewidth]{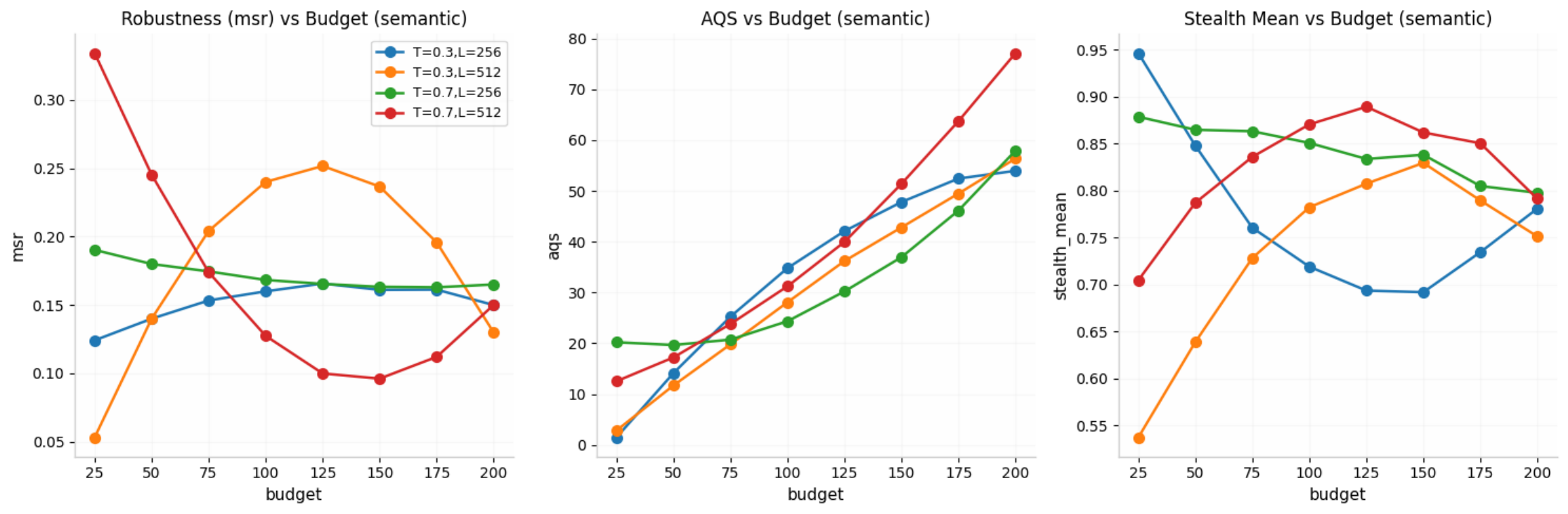}
    \caption{Performance dynamics of single semantic-space attacks across query budgets under four $(T, L)$ configurations: (a) Misuse success rate, (b) Average Queries to Success, and (c) Stealth score.}
    \label{fig.semantic_attack1}
\end{figure}

Semantic-space attack performance shows clear sensitivity to the decoding configuration. As summarized in Table~\ref{tab:semantic_single}, under the conservative setting ($T = 0.3$, $L = 256$), the method achieves a mean MSR of 0.152, a peak MSR of 0.166, a mean AQS of 34.0, and a mean Stealth score of 0.772. Increasing the generation length to $L = 512$ raises the mean and peak MSR to 0.181 and 0.252, respectively, while reducing AQS to 30.9; however, the mean Stealth score decreases to 0.733. This pattern suggests that longer outputs enlarge the semantic search space and can improve attackability, but they also tend to introduce more visible semantic deviation. By contrast, the high-temperature short-sequence setting ($T = 0.7$, $L = 256$) yields the highest mean Stealth score of 0.841, while maintaining only moderate attack success (mean MSR 0.171, peak MSR 0.190), indicating that this configuration favors imperceptibility more than attack strength.

The budget-wise curves in Fig.~\ref{fig.semantic_attack1} further show that semantic-only attacks often obtain their strongest gains early and then become difficult to sustain. In particular, the setting ($T = 0.7$, $L = 512$) reaches a peak MSR of 0.334 at low query budgets and then declines rapidly, forming a clear early-peak pattern. More generally, several configurations exhibit plateauing or decreasing MSR as the budget increases, rather than steady improvement. For example, ($T = 0.3$, $L = 256$) reaches its peak at budget $= 125$ and then declines, while ($T = 0.7$, $L = 512$) drops below 0.10 during the middle budget range. A plausible explanation is that semantic-only mutation can exhaust effective reformulations relatively quickly; after the most useful variants have been explored, further mutation may introduce semantic drift rather than improving attack effectiveness.

Overall, the semantic space provides relatively favorable imperceptibility, but its gains in attack success are limited and often unstable across larger budgets. The results also show a visible trade-off between success rate and Stealth: configurations with higher attack success tend to sacrifice concealment, whereas the most stealthy setting does not achieve the strongest attack performance. Within the tested configurations, semantic-only mutation therefore appears better suited to generating natural-looking variants than to sustaining strong attack effectiveness over long search horizons.

\begin{table*}[htbp]
    \centering
    \caption{Summary of attack performance metrics under single 
             semantic-space mutation across four $(T, L)$ configurations.}
    \label{tab:semantic_single}
    \begin{tabular}{lcccccc}
        \toprule
        \textbf{Config} & \textbf{Mean MSR} & \textbf{Mean AQS} 
        & \textbf{Mean Stealth} & \textbf{Peak MSR} 
        & \textbf{Best AQS} & \textbf{Peak Stealth} \\  
        \midrule
        $T$ = 0.3, $L$ = 256 & 0.152 & 34.0 & 0.772 & 0.166 & 1.5  & 0.946 \\
        $T$ = 0.3, $L$ = 512 & 0.181 & 30.9 & 0.733 & 0.252 & 2.8  & 0.807 \\
        $T$ = 0.7, $L$ = 256 & 0.171 & 32.0 & 0.841 & 0.190 & 20.2 & 0.879 \\
        $T$ = 0.7, $L$ = 512 & 0.167 & 39.6 & 0.824 & 0.334 & 12.5 & 0.889 \\
        \bottomrule
    \end{tabular}
\end{table*}

\subsubsection{Character-Space Attack Performance Analysis}

\begin{figure}[htbp]
    \centering
    \includegraphics[width=1\linewidth]{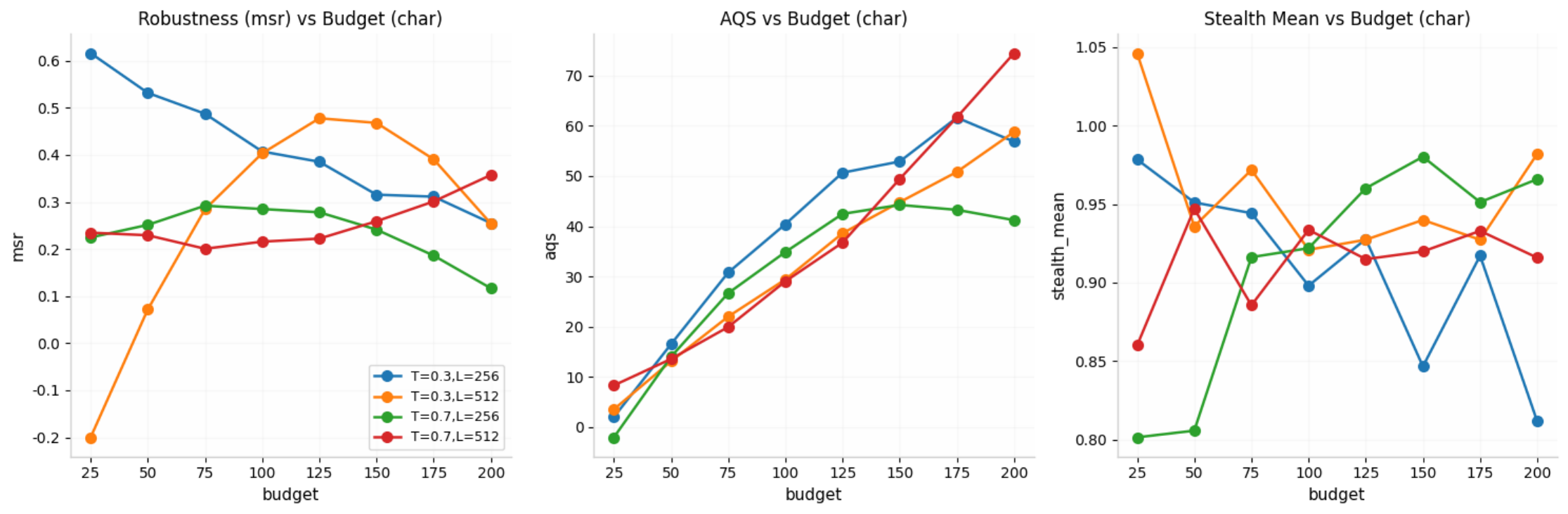}
    \caption{Performance dynamics of single character-space attacks across query budgets under four $(T, L)$ configurations: (a) Misuse success rate, (b) Average Queries to Success, and (c) Stealth score.}
    \label{fig.character_attack1}
\end{figure}

\begin{table*}[htbp]
    \centering
    \caption{Summary of attack performance metrics under single 
             character-space mutation across four $(T, L)$ configurations.}
    \label{tab:Character_single}
    \begin{tabular}{lcccccc}
        \toprule
        \textbf{Config} & \textbf{Mean MSR} & \textbf{Mean AQS} 
        & \textbf{Mean Stealth} & \textbf{Peak MSR} 
        & \textbf{Best AQS} & \textbf{Peak Stealth} \\  
        \midrule
        $T$ = 0.3, $L$ = 256 & 0.248 & 39.0 & 0.818 & 0.370 & 16.5  & 0.978 \\
        $T$ = 0.3, $L$ = 512 & 0.177 & 32.6 & 0.861 & 0.287 & 3.6  & 0.982 \\
        $T$ = 0.7, $L$ = 256 & 0.141 & 30.9 & 0.822 & 0.175 & 14.0 & 0.882 \\
        $T$ = 0.7, $L$ = 512 & 0.151 & 36.7 & 0.822 & 0.214 & 8.3 & 0.966 \\
        \bottomrule
    \end{tabular}
\end{table*}

Character-space mutation performs best under the low-temperature short-sequence setting ($T = 0.3$, $L = 256$), as shown in Table~\ref{tab:Character_single}. Under this configuration, the method achieves a mean MSR of 0.248, a peak MSR of 0.370, a mean AQS of 39.0, and a mean Stealth score of 0.818. When the generation length is increased to $L = 512$, the mean and peak MSR decline to 0.177 and 0.287, respectively, while the mean Stealth score increases to 0.861. This indicates that, in the character space, longer generations do not necessarily strengthen the attack; instead, they tend to improve concealment while diluting attack intensity. High-temperature settings are generally less favorable, with mean MSR values dropping to 0.141--0.151, suggesting that character-level mutation is particularly sensitive to sampling randomness. The budget-wise trends further highlight the dynamic characteristics of character-space attacks. Under ($T = 0.3$, $L = 256$), the method exhibits a rapid-rise-rapid-decline pattern: the MSR reaches 0.370 at budget $= 25$, but then falls to 0.153 by budget $= 200$. In contrast, ($T = 0.3$, $L = 512$) shows a more gradual trajectory, rising from a low initial level to a peak of 0.287 at budget $= 125$, followed by a decline at later budgets. These curves suggest that character-level perturbation can produce strong early breakthroughs under favorable settings, but such gains are difficult to maintain as the search continues. At the same time, the relatively high Stealth scores across configurations indicate that character mutation can preserve surface-level concealment even when attack success is less stable.

Overall, character-space mutation is effective for short-budget probing under specific parameter settings, especially at low temperature. However, its performance is more volatile than its Stealth scores alone would suggest, and its success rate degrades noticeably outside the best-performing configuration. Within the tested settings, character-only mutation therefore appears to offer strong early attack bursts, but limited stability and relatively high dependence on configuration choice.

\subsubsection{Dual-Space Attack Performance Analysis}

\begin{figure}[htbp]
    \centering
    \includegraphics[width=1\linewidth]{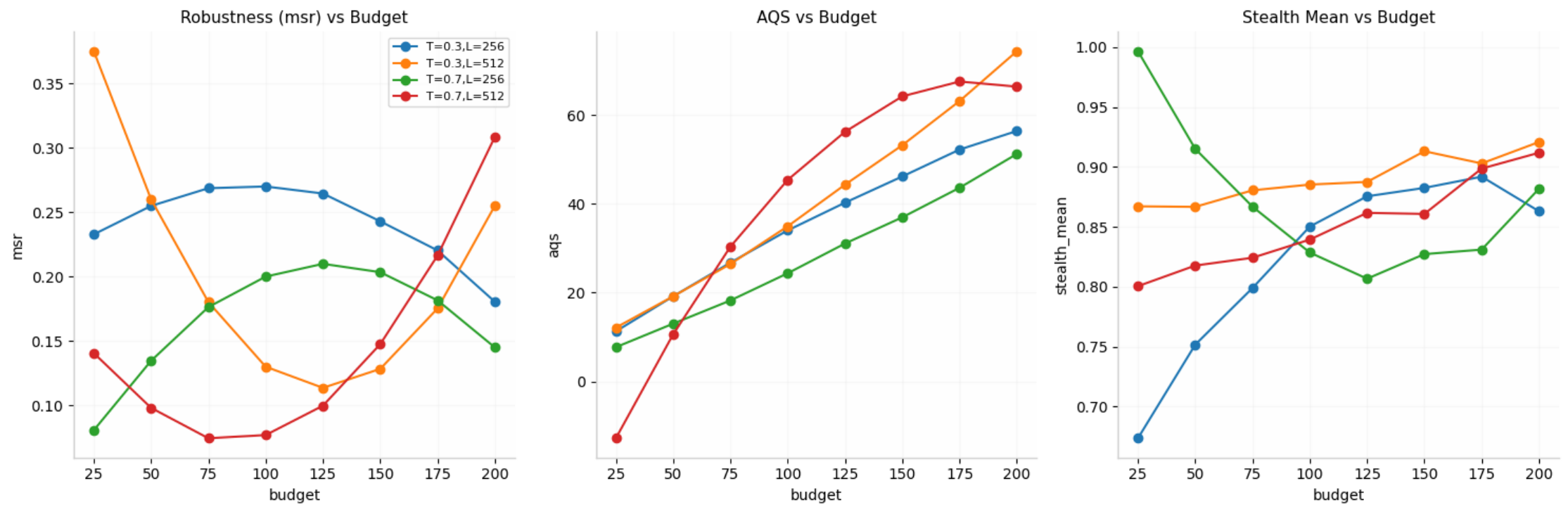}
    \caption{Performance dynamics of dual-space cooperative attacks across query budgets under four $(T, L)$ configurations: (a) Misuse success rate, (b) Average queries to success, and (c) Stealth score.}
    \label{fig.dual_attack1}
\end{figure}

Dual-space mutation shows comparatively balanced behavior across the tested configurations, as summarized in Table~\ref{tab:dual-space} and Fig.~\ref{fig.dual_attack1}. Under the low-temperature short-sequence setting ($T = 0.3$, $L = 256$), the method achieves a mean MSR of 0.242, a peak MSR of 0.270, a mean AQS of 35.8, and a mean Stealth score of 0.823, indicating a relatively even trade-off among the three evaluation dimensions. A different pattern appears under ($T = 0.3$, $L = 512$): although the mean MSR decreases to 0.202, the peak MSR rises sharply to 0.375, while the mean Stealth score also reaches 0.890. This result suggests that the longer low-temperature setting is particularly favorable for discovering high-quality attack instances, even if the average performance becomes more dispersed. Under the high-temperature short-sequence setting ($T = 0.7$, $L = 256$), dual-space mutation records the lowest AQS among its four configurations (28.3), together with a mean MSR of 0.166 and a mean Stealth score of 0.869, indicating stronger query efficiency. The remaining high-temperature long-sequence setting ($T = 0.7$, $L = 512$) shows a lower mean MSR of 0.145 but still reaches a peak MSR of 0.308, suggesting that effective attack instances can still emerge at later stages of the search.

The budget-wise curves reveal two notable dynamic patterns. First, under ($T = 0.3$, $L = 256$), the attack rapidly rises to a near-peak level and then remains in a relatively stable plateau over a substantial budget interval, rather than showing an immediate collapse after early success. Second, under ($T = 0.7$, $L = 512$), the curve displays a late-stage breakthrough pattern, reaching its strongest performance near the end of the tested budget range. Compared with the sharper rise-and-fall behavior observed in single-space settings, these trajectories suggest that dual-space search is less brittle under budget expansion and can maintain useful search momentum across a wider range of query budgets.

A further observation is that successful dual-space samples tend to involve alternating semantic and character operators, with an approximate semantic-to-character ratio of 3:2. Although this observation does not by itself establish a causal mechanism, it is consistent with the intuition that mixed-space mutation may help avoid premature saturation in either search space. Overall, dual-space mutation appears to provide the most balanced combination of success rate, search efficiency, and imperceptibility among the tested settings, while also exhibiting relatively stable budget-wise behavior.

\begin{table*}[htbp]
    \centering
    \caption{Summary of attack performance metrics under dual-space cooperative mutation across four $(T, L)$ configurations.}
    \label{tab:dual-space}
    \begin{tabular}{lcccccc}
        \toprule
        \textbf{Config} & \textbf{Mean MSR} & \textbf{Mean AQS} 
        & \textbf{Mean Stealth} & \textbf{Peak MSR} 
        & \textbf{Best AQS} & \textbf{Peak Stealth} \\  
        \midrule
        $T$ = 0.3, $L$ = 256 & 0.242 & 35.8 & 0.823 & 0.270 & 11.3  & 0.850 \\
        $T$ = 0.3, $L$ = 512 & 0.202 & 40.9 & 0.890 & 0.375 & 12.1  & 0.921 \\
        $T$ = 0.7, $L$ = 256 & 0.166 & 28.3 & 0.869 & 0.210 & 7.7 & 0.997 \\
        $T$ = 0.7, $L$ = 512 & 0.145 & 41.0 & 0.852 & 0.308 & 12.5 & 0.912 \\
        \bottomrule
    \end{tabular}
\end{table*}

\subsection{Comparative Analysis}
\subsubsection{Three-Dimensional Quantitative Comparison}

\begin{figure}[htbp]
    \centering
    \includegraphics[width=1\linewidth]{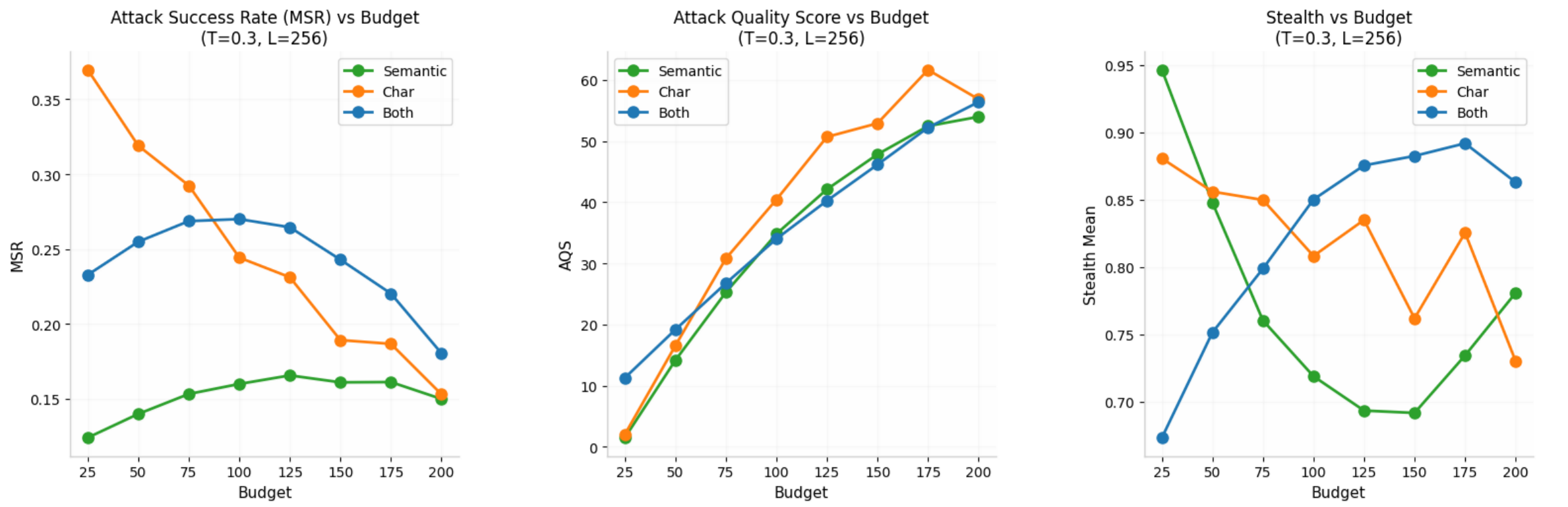}
    \caption{Three-dimensional comparative performance of semantic-only, character-only, and dual-space strategies under $T=0.3$, $L=256$ configuration across query budgets: (a) MSR, (b) AQS, and (c) Stealth.}
    \label{fig.compare1}
\end{figure}

\begin{table*}[htbp]
    \centering
    \footnotesize
    \caption{Three-dimensional quantitative comparison of semantic-only, character-only, and dual-space mutation strategies across the attack success rate, search efficiency, and imperceptibility.}
    \label{tab:three_dim_comparison}
    \begin{tabular}{lcccc}
        \toprule
        \textbf{Evaluation Dimension} 
        & \textbf{Semantic-only} 
        & \textbf{Character-only} 
        & \textbf{Dual-space} 
        & \textbf{Dual- vs.\ Single-Space} \\
        \midrule
        Attack Success Rate (Mean MSR)   & 0.168   & 0.179   & 0.189   & +12.5\% / +5.6\%    \\
        Attack Success Rate (Peak MSR)   & 0.334   & 0.370   & 0.375   & +12.3\% / +1.4\%    \\
        Search Efficiency (Mean AQS)     & 34.1    & 34.8    & 36.5    & +7.0\% / $+$4.9\%   \\
        Search Efficiency (typical AQS)  & 30.9    & 30.9    & 28.3    & $-$8.4\% / $-$8.4\% \\
        Imperceptibility (Mean Stealth)  & 0.793   & 0.831   & 0.859   & +8.3\% / +3.4\%     \\
        Imperceptibility (Peak Stealth)  & 0.946   & 0.982   & 0.997   & +5.4\% / $+$1.5\%   \\
        Per-Query Productivity (MSR/AQS) & 0.00492 & 0.00513 & 0.00518 & +5.3\% / +1.0\%     \\
        \bottomrule
    \end{tabular}
\end{table*}

Based on the results of the three mutation strategies, this section provides a cross-strategy comparison from three complementary perspectives: attack success rate, search efficiency, and imperceptibility.

From the perspective of attack success rate, dual-space mutation shows the strongest aggregate performance among the tested strategies. As reported in Table~\ref{tab:three_dim_comparison}, it achieves the highest mean MSR (0.189) and the highest peak MSR (0.375), indicating that it performs strongly both on average and at its best observed operating point. Character-space mutation remains competitive in this dimension, especially in its strongest configuration, where it reaches a peak MSR close to the dual-space result. By contrast, semantic-space mutation shows a lower overall success rate, although it can still produce relatively strong peak behavior under specific settings. Taken together, these results indicate that dual-space mutation provides the strongest overall attack capability across the tested settings, while the two single-space strategies depend more heavily on favorable parameter choices.

In terms of search efficiency, the relationship among the three strategies is more nuanced. Since lower AQS indicates that successful attacks are found with fewer queries, the best observed efficiency is achieved by the dual-space setting with $T = 0.7$ and $L = 256$, which reaches an AQS of 28.3. At the aggregated level, however, dual-space mutation does not achieve the lowest mean AQS: its mean AQS is 36.5, compared with 34.1 for semantic-only and 34.8 for character-only. This indicates that the efficiency advantage of dual-space mutation is concentrated in its best configuration rather than uniformly maintained across all settings. For this reason, Table~\ref{tab:three_dim_comparison} also reports per-query productivity (MSR/AQS), under which dual-space mutation remains slightly better than the two single-space baselines. Overall, these results suggest that dual-space mutation improves attack success while preserving competitive best-case query efficiency, although its average AQS is slightly higher than those of the single-space baselines.

The comparison in imperceptibility further highlights the advantage of dual-space mutation. It achieves the highest mean Stealth score among the three strategies, indicating that its generated attacks are, on average, less distinguishable from the original prompts under the adopted metric. Character-space mutation can still achieve very high peak Stealth in individual configurations, but this advantage is not consistently accompanied by equally strong success rates. Semantic-space mutation shows moderate concealment overall, but its stronger attack configurations tend to be associated with lower Stealth scores. In contrast, dual-space mutation maintains relatively strong Stealth while also preserving the highest observed attack ceiling, suggesting a more favorable balance between concealment and effectiveness.

Overall, the three-dimensional comparison shows clear differences in strategic profile. Semantic-space mutation is comparatively conservative, with reasonable imperceptibility but limited attack gains. Character-space mutation can produce strong early breakthroughs and high concealment in certain settings, but its performance is more configuration-sensitive. Dual-space mutation provides the most balanced behavior across the three dimensions under the tested settings, making it the strongest overall strategy in this study.

\subsubsection{Parameter Sensitivity Comparison}

\begin{figure}[htbp]
    \centering
    \includegraphics[width=1\linewidth]{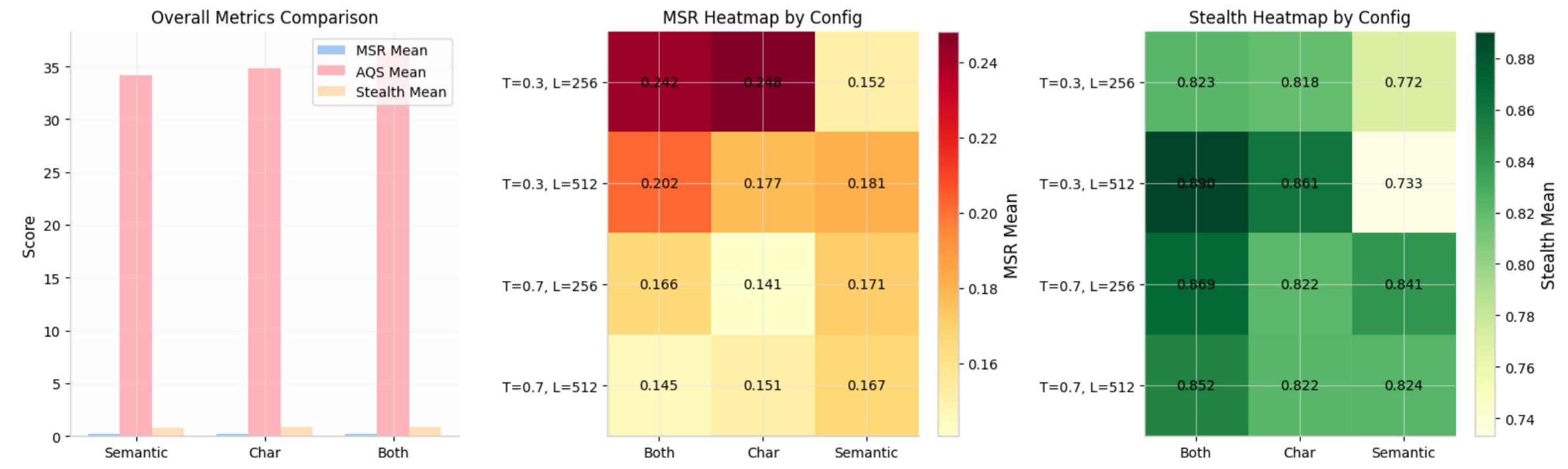}
    \caption{Parameter sensitivity comparison across mutation strategies: 
             (a) overall metric summary (MSR mean, AQS mean, and Stealth mean), 
             (b) MSR heatmap by $(T, L)$ configuration, and 
             (c) Stealth heatmap by $(T, L)$ configuration.}
    \label{fig:sensitivity1}
\end{figure}

\begin{table*}[htbp]
    \centering
    \footnotesize
    \caption{Consolidated performance summary across all mutation strategies and representative $(T, L)$
             configurations, including characteristic behavioral labels.}
    \label{tab:param_sensitivity}
    \setlength{\tabcolsep}{4pt}
    \begin{tabular}{llccccccc}
        \toprule
        \textbf{Strategy} & \textbf{Config}
        & \textbf{Mean MSR} & \textbf{Mean AQS} & \textbf{Mean Stealth}
        & \textbf{Peak MSR} & \textbf{Best AQS} & \textbf{Peak Stealth}
        & \textbf{Label} \\
        \midrule
        \multirow{2}{*}{Semantic-only}
        & $T$=0.3, $L$=256 & 0.152 & 34.0 & 0.772 & 0.166 & 1.5  & 0.946
            & Conservative \\
        & $T$=0.7, $L$=512 & 0.167 & 39.6 & 0.824 & 0.334 & 12.5 & 0.889
            & Stochastic early peak \\
        \midrule
        \multirow{3}{*}{Character-only}
        & $T$=0.3, $L$=256 & 0.248 & 39.0 & 0.818 & 0.370 & 16.5 & 0.978
            & Early burst \\
        & $T$=0.3, $L$=512 & 0.177 & 32.6 & 0.861 & 0.287 & 3.6  & 0.982
            & Max.\ imperceptibility \\
        & $T$=0.7, $L$=512 & 0.151 & 36.7 & 0.822 & 0.214 & 8.3  & 0.966
            & Low overall \\
        \midrule
        \multirow{2}{*}{Dual-space}
        & $T$=0.3, $L$=512 & 0.202 & 40.9 & 0.890 & 0.375 & 12.1 & 0.921
            & MSR--Stealth synergy \\
        & $T$=0.7, $L$=256 & 0.166 & 28.3 & 0.869 & 0.210 & 7.7  & 0.997
            & Min.\ AQS \\
        \bottomrule
    \end{tabular}
\end{table*}

As shown in Fig.~\ref{fig:sensitivity1} and Table~\ref{tab:param_sensitivity}, the effects of temperature differ across the three mutation strategies, but the resulting stability pattern is not the same as the overall performance ranking. Under low-temperature ($T = 0.3$) settings, considering the two generation-length configurations, semantic-only exhibits the smallest variation in mean MSR (standard deviation 0.015), followed by dual-space (0.020), while character-only shows the largest fluctuation (0.036). Under high-temperature ($T = 0.7$) settings, semantic-only again remains the most stable (standard deviation 0.002), character-only stays relatively stable (0.005), and dual-space shows the largest variation (0.011). These results suggest that semantic-only is the most stable strategy with respect to parameter changes, whereas dual-space tends to trade greater variability for stronger aggregate attack performance and higher attack ceilings. Character-only lies between the two, with stability that depends on the specific temperature regime.

The effect of the generation-length parameter is clearly non-monotonic and depends on both the mutation strategy and the temperature setting. For semantic-only, under the low-temperature setting ($T = 0.3$), increasing $L$ from 256 to 512 improves mean MSR by 19.1\% (0.152 to 0.181) while reducing mean Stealth by 5.1\% (0.772 to 0.733), indicating a clear success-rate--imperceptibility trade-off. Under the high-temperature setting ($T = 0.7$), however, the same length increase changes mean MSR only slightly (0.171 to 0.167) while reducing mean Stealth from 0.841 to 0.824. For character-only, the pattern differs across temperatures: under $T = 0.3$, increasing $L$ from 256 to 512 decreases mean MSR by 28.6\% (0.248 to 0.177) but improves mean Stealth by 5.3\% (0.818 to 0.861); under $T = 0.7$, the same change slightly improves mean MSR from 0.141 to 0.151 while leaving mean Stealth essentially unchanged (0.822 to 0.822). For dual-space, the low-temperature setting exhibits a particularly interesting separation between average and best-case behavior: increasing $L$ from 256 to 512 decreases mean MSR by 16.5\% (0.242 to 0.202), yet peak MSR rises by 38.9\% (0.270 to 0.375) and mean Stealth improves by 8.1\% (0.823 to 0.890). Under $T = 0.7$, increasing $L$ from 256 to 512 again reduces mean MSR (0.166 to 0.145) but raises peak MSR from 0.210 to 0.308, while mean Stealth declines slightly from 0.869 to 0.852. Overall, these results indicate that the effect of generation length cannot be characterized by a simple linear trend; instead, it interacts strongly with both mutation space and temperature.

The morphology of the budget growth curves reveals further practical implications. Semantic-only generally follows an early-peak pattern, often followed by decline or low-level plateauing, making it suitable for short-horizon exploration but less sustainable over extended budgets. Character-only does not exhibit a single uniform profile: under favorable settings it can produce strong early bursts, while in other settings it shows a more gradual rise followed by later decline. Dual-space, by contrast, operates in two characteristic modes: rapid convergence to a relatively high plateau or delayed deep breakthrough, combining efficiency with depth and adapting naturally to different time constraints. Concretely, the $T = 0.3$, $L = 256$ configuration reaches a near-peak MSR of 0.269 at budget $= 75$ and sustains it through budget $= 125$, offering a 50-unit stable testing window; the $T = 0.7$, $L = 512$ configuration accumulates momentum prior to budget $= 150$ and then surges rapidly, making it well suited to deep-audit scenarios.

\subsubsection{Practical Implications and Configuration Recommendations}

The comparative results suggest that the three mutation strategies are suitable for different testing objectives rather than a single universal deployment setting. Semantic-space mutation provides relatively natural prompt variants and favorable concealment, but its attack gains are limited and often difficult to sustain as the query budget increases. Character-space mutation can produce strong early breakthroughs under favorable settings, especially at low temperature, but its performance is more sensitive to parameter choice and tends to degrade outside its best-performing configuration. Among the three strategies, dual-space mutation shows the most balanced overall behavior across success rate, query efficiency, and Stealth under the tested settings.

For practical evaluation, parameter selection should be guided by the target objective. If the goal is to maximize the observed attack ceiling while maintaining strong concealment, the dual-space setting with $T = 0.3$ and $L = 512$ is the strongest option in the current experiments, achieving the highest peak MSR together with a high Stealth score. If the goal is rapid probing under limited query resources, the dual-space setting with $T = 0.7$ and $L = 256$ is more suitable because it yields the lowest observed AQS. For evaluations emphasizing relatively stable behavior across budget growth, the dual-space setting with $T = 0.3$ and $L = 256$ provides a more sustained plateau than the more volatile single-space settings. In contrast, semantic-only and character-only mutations appear more suitable as complementary options when a tester specifically wants to emphasize semantic reformulation or shallow obfuscation behavior.

The results also suggest that budget allocation should depend on the mutation space. Semantic-space mutation tends to produce useful variants early, but extended search often yields diminishing returns and may introduce semantic drift. Character-space mutation can be effective for short-budget probing, yet its gains are more configuration-sensitive and less stable over longer search horizons. Dual-space mutation remains the most flexible strategy across the full budget range and therefore serves as the most practical default option in the current framework, especially when prior knowledge of the target defense behavior is limited. Overall, the findings indicate that effective attack evaluation benefits from adaptive configuration selection rather than reliance on a single fixed strategy.

\subsection{Discussion on Defense Implications}

The experimental observations in this study provide several implications for defense design. 

First, the differing behavior of semantic-space, character-space, and dual-space mutations suggests that single-layer defenses are unlikely to be sufficient. Semantic reformulation may weaken rigid lexical filtering, while character-level perturbation can obscure surface-form patterns. When these two mutation modes are combined, the resulting prompt variants become more diverse and less predictable. This suggests that robust defense should rely on multiple complementary detection mechanisms rather than a single rule-based or model-based component.

Second, the results suggest that defense models may benefit from jointly modeling semantic and character-level information. In the current experiments, semantic-only attacks tend to preserve fluency while altering intent expression, whereas character-only attacks preserve much of the surface structure while perturbing textual form. Dual-space attacks combine both effects, which may create blind spots if the defense pipeline evaluates these dimensions separately. A plausible implication is that defense systems should learn cross-level representations capable of identifying prompts that appear benign in one feature space but abnormal in another \cite{guo2019certified}.

Third, the strategy-level results also suggest a rough correspondence between defense depth and attack-space effectiveness. For relatively shallow defenses, such as lexical filtering or simple statistical checks, character-space mutation may already provide meaningful probing capability at relatively low cost. As defenses rely more heavily on semantic understanding or intent recognition, however, single-space mutation appears less reliable, and mixed semantic-character search becomes comparatively more useful. At the deepest level, where behavior is constrained by stronger alignment mechanisms, all three strategies show bounded effectiveness. This indicates that deeper safety alignment remains an important barrier even when shallow defenses are partially bypassed.

Fourth, the budget-wise attack dynamics indicate that static responses may unintentionally provide optimization signals to adaptive attackers. In particular, iterative search-based attacks can exploit relatively stable feedback patterns to refine prompts over time. This may help explain why some configurations achieve rapid early gains or late-stage breakthroughs under continued querying. From the defense perspective, dynamic response strategies, such as response randomization, delayed feedback, or uncertainty-aware moderation, may reduce the usefulness of attacker-side feedback loops and make adaptive optimization less effective \cite{chen2018detecting}.

These observations also carry implications for evaluation practice. Because attack effectiveness varies with both defense depth and query budget, security testing may benefit from adaptive strategy selection rather than a fixed single-space setup. For example, early probing results can be used to determine whether shallow obfuscation is already sufficient or whether broader dual-space exploration is needed. Such adaptive evaluation may provide a more realistic estimate of system robustness than relying on only one mutation mode or one budget regime. Overall, the findings support a defense strategy that combines heterogeneous detection, joint representation learning, dynamic response design, and continued alignment reinforcement \cite{carlini2021extracting}. At the same time, they suggest that robustness evaluation should account for the interaction between defense depth, mutation space, and query budget, rather than treating prompt attacks as a single uniform threat model.

\section{Conclusion and Future work} \label{sec: conclusion}

This paper presents PromptFuzz-SC, a dual-space adversarial prompt generation framework for prompt-level security evaluation of large language models. By jointly modeling semantic and character mutations within a unified search framework, the proposed method expands the adversarial search space and enables more comprehensive exploration of model vulnerabilities beyond single-space strategies. Experimental results demonstrate that dual-space mutation achieves the strongest overall performance among the evaluated settings, reaching the highest mean MSR (0.189), peak MSR (0.375), and mean Stealth (0.859). Compared with semantic-only and character-only mutation, it improves mean MSR by 12.5\% and 5.6\%, respectively, with corresponding peak MSR gains of 12.3\% and 1.4\%. Under specific configurations, the method attains a peak MSR of 0.375 while maintaining high Stealth (0.890), indicating a favorable balance between attack effectiveness and imperceptibility. Although it does not consistently minimize query cost, dual-space mutation achieves competitive best-case efficiency (AQS = 28.3) and exhibits flexible attack patterns across different settings. These results highlight the practical value of composite mutation strategies for red-team testing against multi-layer LLM defense pipelines.

Despite these contributions, several limitations remain. The current evaluation is restricted to a general-purpose LLM setting and does not cover domain-specific deployments or specialized defense mechanisms. Moreover, the query budget is limited to 200, constraining the analysis of long-horizon adaptive attacks. Future work will extend this framework toward multimodal adversarial settings, adaptive attack–defense co-evolution, and more efficient automated search strategies. In addition, developing standardized cross-domain benchmarks and integrating stronger safety safeguards will be essential for advancing robust and responsible LLM security evaluation.

\bibliographystyle{IEEEtran}
\bibliography{main}

@inproceedings{gan2023model,
  title={Model-as-a-service ({MaaS}): A survey},
  author={Gan, Wensheng and Wan, Shicheng and Yu, Phillip S},
  booktitle={IEEE International Conference on Big Data},
  pages={4636--4645},
  year={2023},
  organization={IEEE}
}

@inproceedings{wu2023multimodal,
  title={Multimodal large language models: A survey},
  author={Wu, Jiayang and Gan, Wensheng and Chen, Zefeng and Wan, Shicheng and Yu, Phillip S},
  booktitle={IEEE International Conference on Big Data},
  pages={2247--2256},
  year={2023},
  organization={IEEE}
}

@article{gan2025mixture,
  title={Mixture of experts ({MoE}): A big data perspective},
  author={Gan, Wensheng and Ning, Zhenyao and Qi, Zhenlian and Yu, Philip S},
  journal={Information Fusion},
  pages={1--28},
  year={2025},
  publisher={Elsevier}
}

@article{wang2025safety,
  title={Safety in large reasoning models: A survey},
  author={Wang, Cheng and Liu, Yue and Li, Baolong and Zhang, Duzhen and Li, Zhongzhi and Fang, Junfeng},
  journal={arXiv preprint arXiv:2504.17704},
  year={2025}
}

@inproceedings{chao2025jailbreaking,
  title={Jailbreaking black box large language models in twenty queries},
  author={Chao, Patrick and Robey, Alexander and Dobriban, Edgar and Hassani, Hamed and Pappas, George J and Wong, Eric},
  booktitle={IEEE Conference on Secure and Trustworthy Machine Learning},
  pages={23--42},
  year={2025},
  organization={IEEE}
}

@inproceedings{greshake2023not,
  title={Not what you've signed up for: Compromising real-world {LLM}-integrated applications with indirect prompt injection},
  author={Greshake, Kai and Abdelnabi, Sahar and Mishra, Shailesh and Endres, Christoph and Holz, Thorsten and Fritz, Mario},
  booktitle={The ACM Workshop on Artificial Intelligence and Security},
  pages={79--90},
  year={2023}
}

@article{paulus2024advprompter,
  title={{AdvPrompter}: Fast adaptive adversarial prompting for {LLMs}},
  author={Paulus, Anselm and Zharmagambetov, Arman and Guo, Chuan and Amos, Brandon and Tian, Yuandong},
  journal={arXiv preprint arXiv:2404.16873},
  year={2024}
}

@article{steinhardt2017certified,
  title={Certified defenses for data poisoning attacks},
  author={Steinhardt, Jacob and Koh, Pang Wei W and Liang, Percy S},
  journal={Advances in Neural Information Processing Systems},
  volume={30},
  year={2017}
}

@inproceedings{zhang2024defending,
  title={Defending large language models against jailbreaking attacks through goal prioritization},
  author={Zhang, Zhexin and Yang, Junxiao and Ke, Pei and Mi, Fei and Wang, Hongning and Huang, Minlie},
  booktitle={The Annual Meeting of the Association for Computational Linguistics},
  pages={8865--8887},
  year={2024}
}

@article{anil2024many,
  title={Many-shot jailbreaking},
  author={Anil, Cem and Durmus, Esin and Panickssery, Nina and Sharma, Mrinank and Benton, Joe and Kundu, Sandipan and Batson, Joshua and Tong, Meg and Mu, Jesse and Ford, Daniel and others},
  journal={Advances in Neural Information Processing Systems},
  volume={37},
  pages={129696--129742},
  year={2024}
}

@inproceedings{wan2023poisoning,
  title={Poisoning language models during instruction tuning},
  author={Wan, Alexander and Wallace, Eric and Shen, Sheng and Klein, Dan},
  booktitle={International Conference on Machine Learning},
  pages={35413--35425},
  year={2023}
}

@inproceedings{wallace2019universal,
  title={Universal adversarial triggers for attacking and analyzing NLP},
  author={Wallace, Eric and Feng, Shi and Kandpal, Nikhil and Gardner, Matt and Singh, Sameer},
  booktitle={The Conference on EMNLP-IJCNLP},
  pages={2153--2162},
  year={2019}
}

@inproceedings{shin2020autoprompt,
  title={{AutoPrompt}: Eliciting knowledge from language models with automatically generated prompts},
  author={Shin, Taylor and Razeghi, Yasaman and Logan IV, Robert L and Wallace, Eric and Singh, Sameer},
  booktitle={The Conference on Empirical Methods in Natural Language Processing},
  pages={4222--4235},
  year={2020}
}

@article{zou2023universal,
  title={Universal and transferable adversarial attacks on aligned language models},
  author={Zou, Andy and Wang, Zifan and Carlini, Nicholas and Nasr, Milad and Kolter, J Zico and Fredrikson, Matt},
  journal={arXiv preprint arXiv:2307.15043},
  year={2023}
}

@inproceedings{zhu2023promptrobust,
  title={PromptRobust: Towards evaluating the robustness of large language models on adversarial prompts},
  author={Zhu, Kaijie and Wang, Jindong and Zhou, Jiaheng and Wang, Zichen and Chen, Hao and Wang, Yidong and Yang, Linyi and Ye, Wei and Zhang, Yue and Gong, Neil and others},
  booktitle={The 1st ACM workshop on Large AI Systems and Models with Privacy and Safety Analysis},
  pages={57--68},
  year={2023}
}

@inproceedings{li2024red,
  title={Red teaming visual language models},
  author={Li, Mukai and Li, Lei and Yin, Yuwei and Ahmed, Masood and Liu, Zhenguang and Liu, Qi},
  booktitle={Findings of the Association for Computational Linguistics},
  pages={3326--3342},
  year={2024}
}

@article{cheng2025exploring,
  title={Exploring typographic visual prompts injection threats in cross-modality generation models},
  author={Cheng, Hao and Xiao, Erjia and Wang, Yichi and Zhang, Lingfeng and Zhang, Qiang and Cao, Jiahang and Xu, Kaidi and Sun, Mengshu and Hao, Xiaoshuai and Gu, Jindong and others},
  journal={arXiv preprint arXiv:2503.11519},
  year={2025}
}

@article{nasr2025attacker,
  title={The attacker moves second: Stronger adaptive attacks bypass defenses against {LLM} jailbreaks and prompt injections},
  author={Nasr, Milad and Carlini, Nicholas and Sitawarin, Chawin and Schulhoff, Sander V and Hayes, Jamie and Ilie, Michael and Pluto, Juliette and Song, Shuang and Chaudhari, Harsh and Shumailov, Ilia and others},
  journal={arXiv preprint arXiv:2510.09023},
  year={2025}
}

@article{ni2025shieldlearner,
  title={Shieldlearner: A new paradigm for jailbreak attack defense in {LLMs}},
  author={Ni, Ziyi and Wang, Hao and Wang, Huacan},
  journal={arXiv preprint arXiv:2502.13162},
  year={2025}
}

@inproceedings{zhou2025hidden,
  title={The hidden risks of large reasoning models: A safety assessment of {R1}},
  author={Zhou, Kaiwen and Liu, Chengzhi and Zhao, Xuandong and Jangam, Shreedhar and Srinivasa, Jayanth and Liu, Gaowen and Song, Dawn and Wang, Xin Eric},
  booktitle={The 14th International Joint Conference on Natural Language Processing and the 4th Conference of the Asia-Pacific Chapter of the Association for Computational Linguistics},
  pages={3250--3265},
  year={2025}
}

@inproceedings{wu2025sugar,
  title={Sugar-Coated Poison: Benign Generation Unlocks Jailbreaking},
  author={Wu, Yu-Hang and Xiong, Yu-Jie and Zhang, Hao and Zhang, Jia-Chen and Zhou, Zheng},
  booktitle={Findings of the Association for Computational Linguistics},
  pages={9645--9665},
  year={2025}
}

@article{cui2025token,
  title={Token-efficient prompt injection attack: Provoking cessation in {LLM} reasoning via adaptive token compression},
  author={Cui, Yu and Cai, Yujun and Wang, Yiwei},
  journal={arXiv preprint arXiv:2504.20493},
  year={2025}
}

@article{souly2025poisoning,
  title={Poisoning attacks on {LLMs} require a near-constant number of poison samples},
  author={Souly, Alexandra and Rando, Javier and Chapman, Ed and Davies, Xander and Hasircioglu, Burak and Shereen, Ezzeldin and Mougan, Carlos and Mavroudis, Vasilios and Jones, Erik and Hicks, Chris and others},
  journal={arXiv preprint arXiv:2510.07192},
  year={2025}
}

@article{shafahi2019adversarial,
  title={Adversarial training for free!},
  author={Shafahi, Ali and Najibi, Mahyar and Ghiasi, Mohammad Amin and Xu, Zheng and Dickerson, John and Studer, Christoph and Davis, Larry S and Taylor, Gavin and Goldstein, Tom},
  journal={Advances in neural information processing systems},
  volume={32},
  year={2019}
}

@inproceedings{graves2021amnesiac,
  title={Amnesiac machine learning},
  author={Graves, Laura and Nagisetty, Vineel and Ganesh, Vijay},
  booktitle={The AAAI Conference on Artificial Intelligence},
  volume={35},
  number={13},
  pages={11516--11524},
  year={2021}
}

@inproceedings{shokri2017membership,
  title={Membership inference attacks against machine learning models},
  author={Shokri, Reza and Stronati, Marco and Song, Congzheng and Shmatikov, Vitaly},
  booktitle={IEEE Symposium on Security and Privacy},
  pages={3--18},
  year={2017},
  organization={IEEE}
}

@article{ackerman2025mitigating,
  title={Mitigating many-shot jailbreaking},
  author={Ackerman, Christopher M and Panickssery, Nina},
  journal={arXiv preprint arXiv:2504.09604},
  year={2025}
}

@inproceedings{wang2025selfdefend,
  title={{SelfDefend}: {LLMs} can defend themselves against jailbreaking in a practical manner},
  author={Wang, Xunguang and Wu, Daoyuan and Ji, Zhenlan and Li, Zongjie and Ma, Pingchuan and Wang, Shuai and Li, Yingjiu and Liu, Yang and Liu, Ning and Rahmel, Juergen},
  booktitle={The 34th USENIX Security Symposium},
  pages={2441--2460},
  year={2025}
}

@inproceedings{perez2022red,
  title={Red teaming language models with language models},
  author={Perez, Ethan and Huang, Saffron and Song, Francis and Cai, Trevor and Ring, Roman and Aslanides, John and Glaese, Amelia and McAleese, Nat and Irving, Geoffrey},
  booktitle={The Conference on Empirical Methods in Natural Language Processing},
  pages={3419--3448},
  year={2022}
}

@inproceedings{gao2019strip,
  title={Strip: A defence against trojan attacks on deep neural networks},
  author={Gao, Yansong and Xu, Change and Wang, Derui and Chen, Shiping and Ranasinghe, Damith C and Nepal, Surya},
  booktitle={The 35th Annual Computer Security Applications Conference},
  pages={113--125},
  year={2019}
}

@article{varshney2023stitch,
  title={A stitch in time saves nine: Detecting and mitigating hallucinations of {LLMs} by validating low-confidence generation},
  author={Varshney, Neeraj and Yao, Wenlin and Zhang, Hongming and Chen, Jianshu and Yu, Dong},
  journal={arXiv preprint arXiv:2307.03987},
  year={2023}
}

@article{wang2023decodingtrust,
  title={{DecodingTrust}: A Comprehensive Assessment of Trustworthiness in {GPT} Models},
  author={Wang, Boxin and Chen, Weixin and Pei, Hengzhi and Xie, Chulin and Kang, Mintong and Zhang, Chenhui and Xu, Chejian and Xiong, Zidi and Dutta, Ritik and Schaeffer, Rylan and others},
  journal={Advances in Neural Information Processing Systems},
  volume={36},
  pages={31232--31339},
  year={2023}
}

@article{wei2024emoji,
  title={Emoji attack: Enhancing jailbreak attacks against judge {LLM} detection},
  author={Wei, Zhipeng and Liu, Yuqi and Erichson, N Benjamin},
  journal={arXiv preprint arXiv:2411.01077},
  year={2024}
}

@article{guo2019certified,
  title={Certified data removal from machine learning models},
  author={Guo, Chuan and Goldstein, Tom and Hannun, Awni and Van Der Maaten, Laurens},
  journal={arXiv preprint arXiv:1911.03030},
  year={2019}
}

@article{chen2018detecting,
  title={Detecting backdoor attacks on deep neural networks by activation clustering},
  author={Chen, Bryant and Carvalho, Wilka and Baracaldo, Nathalie and Ludwig, Heiko and Edwards, Benjamin and Lee, Taesung and Molloy, Ian and Srivastava, Biplav},
  journal={arXiv preprint arXiv:1811.03728},
  year={2018}
}

@inproceedings{carlini2021extracting,
  title={Extracting training data from large language models},
  author={Carlini, Nicholas and Tramer, Florian and Wallace, Eric and Jagielski, Matthew and Herbert-Voss, Ariel and Lee, Katherine and Roberts, Adam and Brown, Tom and Song, Dawn and Erlingsson, Ulfar and others},
  booktitle={The 30th USENIX Security Symposium},
  pages={2633--2650},
  year={2021}
}

@inproceedings{dong2024survey,
  title={A Survey on In-context Learning},
  author={Dong, Qingxiu and Li, Lei and Dai, Damai and Zheng, Ce and Ma, Jingyuan and Li, Rui and Xia, Heming and Xu, Jingjing and Wu, Zhiyong and Chang, Baobao and others},
  booktitle={The Conference on Empirical Methods in Natural Language Processing},
  pages={1107--1128},
  year={2024}
}

\end{document}